\documentclass[12pt]{article}

\usepackage{amssymb}

\begin{document}

\begin{titlepage}

\begin{flushright}
\end{flushright}
\vskip 2.5cm

\begin{center}
{\Large \bf Contributions to Pion Decay from Lorentz\\
Violation in the Weak Sector }
\end{center}

\vspace{1ex}

\begin{center}
{\large Brett Altschul\footnote{{\tt baltschu@physics.sc.edu}}}

\vspace{5mm}
{\sl Department of Physics and Astronomy} \\
{\sl University of South Carolina} \\
{\sl Columbia, SC 29208} \\
\end{center}

\vspace{2.5ex}

\medskip

\centerline {\bf Abstract}

\bigskip

Lorentz violation in the weak sector would affect the $\beta$-decay lifetimes of
pions. The decay amplitude may be rendered anisotropic, but only an isotropic
violation of boost invariance can affect the net lifetime in the center of mass
frame. However, since the
rest frames of the pions that produce the NuMI neutrino beam at Fermilab vary with the
rotation of the Earth, it is possible to constrain anisotropic Lorentz violation using
prior analyses of sidereal variations in the event rate at the MINOS near detector.
The resulting bounds on weak-sector Lorentz violation are at the $10^{-4}$ level,
a substantial improvement over previous results. The highly relativistic character
of the pions involved is responsible for the improvement.

\bigskip

\end{titlepage}

\newpage

\section{Introduction}

There has recently been a significant degree of interest in the possibility that
the fundamental Lorentz and CPT symmetries of the standard model might not hold
exactly. There has not so far been any significant experimental indication that these
symmetries are broken in nature. However, the possibility is interesting, because if
such symmetry breaking did exist, it would be indicative of a completely new regime of
fundamental physics.

Lorentz and CPT violation involving standard model quanta can be described using
the machinery of effective field theory. The general field theory that describes
these effects is known as the standard model extension (SME). It contains all
possible  translation-invariant but Lorentz-violating operators that may be
constructed out of known standard model fields.
The effects of the Lorentz-violating operators are parameterized by tensor-valued
coefficients~\cite{ref-kost1,ref-kost2}. If Lorentz symmetry is broken spontaneously,
these are related to the vacuum expectation values of
dynamical fields with tensor indices.

The minimal SME---a restricted version of the theory that contains only
renormalizable operators---is now the standard framework used for describing the
results of experimental tests of isotropy, boost invariance, and CPT symmetry.
Important tests of these symmetries have included studies of matter-antimatter
asymmetries for trapped charged
particles~\cite{ref-bluhm1,ref-gabirelse,ref-dehmelt1} and bound state
systems~\cite{ref-bluhm3,ref-phillips},
measurements of muon properties~\cite{ref-kost8,ref-hughes}, analyses of
the behavior of spin-polarized matter~\cite{ref-heckel3},
frequency standard comparisons~\cite{ref-berglund,ref-kost6,ref-bear,ref-wolf},
Michelson-Morley experiments with cryogenic
resonators~\cite{ref-muller3,ref-herrmann3,ref-eisele}, measurements of neutral meson
oscillations~\cite{ref-kost7,ref-hsiung,ref-abe,
ref-link,ref-aubert}, polarization measurements on the light from cosmological
sources~\cite{ref-kost21,ref-kost22},
high-energy astrophysical
tests~\cite{ref-stecker,ref-jacobson1,ref-altschul6,ref-klinkhamer2},
precision tests of gravity~\cite{ref-battat,ref-muller4}, and others.
Up-to-date information about the status of the relevant
constraints may be found collected in~\cite{ref-tables}.

Motivated by the claim from OPERA~\cite{ref-opera} that neutrinos might have been
moving faster than light, we previously looked at the effects of leptonic Lorentz
violation on the pion decay rate. It turned out that the kind of modification to the
neutrino dispersion relation that would have been necessary to explain the
purported OPERA result would have also led to drastic modifications to the
pion decays that produced the neutrinos and thus to the experiment's beam structure.
More recently, there has been interest in constraining Lorentz violation in the
weak sector using analyses of $\beta$-decays~\cite{ref-noordmans,ref-wilschut}.
Studies of the energetics of isotopes' $\beta$-decay endpoints can also provide
sensitivity to other types of Lorentz violation~\cite{ref-diaz}.

This work is a generalization of our previous
analyses~\cite{ref-altschul32,ref-altschul33} of how Lorentz violation could affect
leptonic meson decays, such as $\pi^{-}\rightarrow\mu^{-}+\bar{\nu}_{\mu}$.
The previous work dealt with Lorentz violation for the
second-generation fermions. Our new results account for the possibility that
the $W$ propagator may also be modified by a Lorentz-violating background tensor
$\chi^{\mu\nu}$. This provides an interesting way to constrain weak sector Lorentz
violation, and it turns out that working with relativistic pions provides significant
improvements in sensitivity relative to studying the decays of stationary
particles.

The outline of this paper is as follows. The decay rate for a pion in the presence of
weak-sector Lorentz violation is evaluated in section~\ref{sec-decay}.
In section~\ref{sec-constr}, this result is combined with existing analyses of MINOS
near detector data to place new constraints on the Lorentz violation coefficients
involved.  Section~\ref{sec-disc} discusses these result in context, including a
critique of some previously claimed constraints on a number of the same coefficients.
An appendix extends the calculations in section~\ref{sec-decay} to show how measuring
particle spins could theoretically provide sensitivity to additinal coefficients,
although this is experimentally impractical.

\section{Lorentz Violation in Pion Decay}

\label{sec-decay}

The present calculations will assume that Lorentz violation only exists in
the pure weak and Higgs sectors. This means the tensor structure of the $W$
propagator will be modified, but
the kinematics of the parent and daughter particles will not. The calculation
may be carried out in the rest frame of the pion, and all the kinematical factor
associated with both the incoming and outgoing particles are unchanged from the
usual case. So only the modified matrix element needs to be calculated,
and its modified form is actually quite simple.
A similar calculation, but assuming a real, traceless $\chi$ that was isotropic in
the laboratory frame, was performed in~\cite{ref-nielsen}.
Naturally, since any real Lorentz
violation is a fairly small effect, we may neglect all effects beyond leading order.
The modifications uncovered in the present calculation may therefore
simply be added to those
found in~\cite{ref-altschul33} if there is relevant Lorentz violation in both the weak
and muon sectors.

The specific form of Lorentz violation we shall consider is a CPT-even modification
of the $W$ propagator. In the Feynman gauge, and in the limit in which the exchanged
momentum is negligible compared to the $W$ boson mass ($|p^{2}|\ll m_{W}^{2}$), the
propagator becomes
\begin{equation}
\label{eq-Wprop}
D^{\mu\nu}(p\rightarrow0)=-i\frac{g^{\mu\nu}+\chi^{\mu\nu}}{-m_{W}^{2}}.
\end{equation}
The complex tensor $\chi^{\mu\nu}$ is an effective construction, related to Lorentz
violation coefficients
that arise in the pure gauge and Higgs boson sectors~\cite{ref-anderson,ref-wilschut}.
In the present limit, with the exchanged $W$ far off shell, $\chi$ takes the form
\begin{equation}
\label{eq-chibreakdown}
\chi^{\mu\nu}=-k_{\phi\phi}^{\mu\nu}-\frac{i}{2g}k_{\phi W}^{\mu\nu}+
\frac{2p_{\alpha}p_{\beta}}{m_{W}^{2}}k_{W}^{\alpha\mu\beta\nu}.
\end{equation}
The tensors $k_{\phi\phi}$, $k_{\phi W}$, and $k_{W}$ describe CPT-even forms of
Lorentz violation in the Lagrangian; they relate to the Higgs sector, to
Higgs-weak mixing, and to the pure $W$ sector, respectively. After spontaneous
symmetry breaking, when the Higgs field acquires a vacuum expectation value,
coupling terms between the scalar field and the $W$ produce operators that are
simply bilinear in the $W$ field and so contribute to the $W$ propagator. There are
additional terms in this propagator related to each of the $k_{\phi\phi}$,
$k_{\phi W}$, and $k_{W}$ tensors, but those appearing in (\ref{eq-chibreakdown})
are the dominant contributions from each of those tensors; further terms are
suppressed by relative powers of $p^{2}/m_{W}^{2}$. In the rest frame of a decaying
pion, where the $W$ momentum is purely timelike, the momentum-dependent $k_{W}$ term
in (\ref{eq-chibreakdown}) becomes $2(m_{\pi}^{2}/m_{W}^{2})k_{W}^{0\mu0\nu}$. More
generally, the structure of the $k_{W}$ contribution to the effective $\chi$ will
depend on the experimental conditions, and so results from different decays will
provide complementary sensitivities. However, since the first-order dependence
of $\chi$ on $k_{W}$ is comparable to higher-order corrections to the
$k_{\phi\phi}$ and $k_{\phi W}$ dependences, we shall mostly treat experimental
results pertaining to $\chi$ as providing constraints on the two-index tensors
$k_{\phi\phi}$ and $k_{\phi W}$. Hermiticity ensures that the
symmetric part of $\chi^{\mu\nu}$ is real and the antisymmetric part is
purely imaginary. Consideration of isotropic, boost-invariance-violating $\chi$ terms
dates back to~\cite{ref-nielsen,ref-huerta}, but the analysis here will include
anisotropic terms as well.

In general,
calculations of scattering cross sections and decay rates for reactions with
Lorentz violation can be tricky. The methods for performing such calculations
were worked out in~\cite{ref-kost5}. Kinematic modifications, which are determined
by the energy-momentum relations for the incoming and outgoing particles, are often
more important than any changes to the matrix element. However, the pion, muon, and
neutrino dispersion relations are all conventional in this case, and so the
calculations simplify considerably. All that is necessary is to compute the
matrix element using the modified propagator (\ref{eq-Wprop}) instead of the
usual one, in the usual tree-level Feynman diagram that describes the pion decay.

In the rest frame of a decaying pion, the decay rate can only depend on isotropic
forms of Lorentz violation---those that only violate Lorentz boost symmetry. The
reason is that the unperturbed
decay is isotropic in the center of mass frame. Effects that break
rotation symmetry will produce a decay amplitude that depends on the momentum
directions of the outgoing particles. However, when the decay rate is integrated
over all angular channels, the anisotropies will cancel out; an increased rate for
a daughter particle to be emitted in one direction will be canceled by a decreased
rate for a decay along a different direction. In a general
theory, it is difficult to
demonstrate this fact explicitly; the angular dependence of the decay amplitude is
quite complicated. However, when the only source of Lorentz
violation for this process is in the weak sector, the calculation of the angular
dependence is simple enough to done explicitly. We shall therefore include all
forms of isotropy breaking in our present calculations, with the expectation that
they will all cancel out in the final expression for the pion decay rate.
The tensor $\chi^{\mu\nu}$ consists of a symmetric real part and an antisymmetric
imaginary part, and since the latter has no isotropic component, it will be expected
not to contribute to the final integrated decay rate.

In our partial
evaluation of the matrix element, any factors related to the pion structure
will be neglected. These factors are common to decays with and without Lorentz
violation. Moreover, in the center of mass frame, the only vector or axial vector
than can be associated with the parent particle is its momentum,
$p_{\pi}=(m_{\pi},0,0,0)$. This simplifies the determination of the
Lorentz-violating modifications considerably. However, the daughter particles do have
spatial momenta, which will complicate the determination of the leptonic part of
the matrix element somewhat. The momenta of the outgoing muon and antineutrino are
$p_{\mu}=([1+\xi^{2}]m_{\pi}/2,[(1-\xi^{2})m_{\pi}/2]\hat{p})$ and
$p_{\nu}=([1-\xi^{2}]m_{\pi}/2,-[(1-\xi^{2})m_{\pi}/2]\hat{p})$, where
$\xi=m_{\mu}/m_{\pi}$ is the muon-pion mass ratio. (The labels $\mu$ and $\nu$ in
these momenta denote the particle type and are not Lorentz indices; to avoid
confusion, other letters will be used to denote Lorentz indices in
expressions involving these vectors.) With no Lorentz violation in the purely leptonic
sector, the expressions for $p_{\mu}$ and $p_{\nu}$ are conventional, and there are
no changes to the kinematics of the decay.

The relevant part of the matrix element is given by
\begin{equation}
{\cal M}\propto
p_{\pi}^{\alpha}\left(\frac{g^{\alpha\beta}+\chi^{\alpha\beta}}{-m_{W}^{2}}
\right)\bar{v}_{\nu}(p_{\nu})\gamma_{\beta}(1-\gamma_{5})u_{\mu}(p_{\mu}).
\end{equation}
Squaring this gives
\begin{equation}
|{\cal M}|^{2}\propto\frac{m_{\pi}^{2}}{m_{W}^{4}}\left[\bar{v}_{\nu}(p_{\nu})
(\gamma_{0}+\chi^{0\beta}\gamma_{\beta})(1-\gamma_{5})u_{\mu}(p_{\mu})
\bar{u}_{\mu}(p_{\mu})(1+\gamma_{5})
(\gamma_{0}+\chi^{*0\gamma}\gamma_{\gamma})v_{\mu}(p_{\mu})\right].
\end{equation}
The sum over the outgoing spins can be evaluated using closure relations. Dropping
the $m_{\pi}$ and $m_{W}$ prefactors, this gives
\begin{eqnarray}
\label{eq-M2}
\sum_{s,s'}|{\cal M}|^{2} & \propto & {\rm tr}\left\{\!\not\!p_{\nu}
(\gamma_{0}+\chi^{0\beta}\gamma_{\beta})(1-\gamma_{5})(\!\not\!p_{\mu}+m_{\mu})
(1+\gamma_{5})(\gamma_{0}+\chi^{*0\gamma}\gamma_{\gamma})
\right\} \\
\label{eq-M2simp}
& = & 2(g^{0\beta}+\chi^{0\beta})(g^{0\gamma}+\chi^{*0\gamma}){\rm tr}\left\{
\!\not\!p_{\nu}\gamma_{\beta}\!\not\!p_{\mu}\gamma_{\gamma}-
\!\not\!p_{\nu}\gamma_{\beta}\gamma_{5}\!\not\!p_{\mu}\gamma_{\gamma}\right\} \\
& = & 8\left\{[2(p_{\mu})_{0}(p_{\nu})_{0}-p_{\mu}\cdot p_{\nu}]\right. \nonumber \\
& & +(\chi^{0\beta}+\chi^{*0\beta})[(p_{\mu})_{0}(p_{\nu})_{\beta}+
(p_{\mu})_{\beta}(p_{\nu})_{0}-(p_{\mu}\cdot p_{\nu})g_{0\beta}] \nonumber\\
& & \left.+ig^{0\gamma}(\chi^{0\beta}-\chi^{*0\beta})
\epsilon_{\alpha\beta\delta\gamma}(p_{\nu})^{\alpha}(p_{\mu})^{\delta}\right\}.
\label{eq-M2final}
\end{eqnarray}

The term proportional to $\epsilon_{\alpha\beta\delta\gamma}$ vanishes. While it
was expected that the imaginary part of $\chi$ would not contribute to the total
decay rate (because of its intrinsic anisotropy), it actually makes no
leading-order contribution to the spin-summed $|{\cal M}|^{2}$, even prior to an
angular integration over $\hat{p}$. The reason is that the
$g^{0\gamma}\epsilon_{\alpha\beta\delta\gamma}$ ensures that there
is no contribution from the timelike components of $p_{\mu}$ and $p_{\nu}$; then
since the spacelike components of these vectors are antiparallel, they also give an
identically vanishing contribution when simultaneously contracted with the Levi-Civita
tensor.

There is, however, a dependence on the imaginary part of $\chi$ in the spin structure
of the cross section. Any such dependence on $\chi^{\mu\nu}-\chi^{*\mu\nu}$ must be
related to muon spin correlations, since the neutrino has only a single interacting
spin
state. The $W$ boson only interacts with a single chirality state of the muon, but
for a massive particle, $\gamma_{5}$ does not commute with the free-particle
Hamiltonian, so both muon spin states are involved in the reaction. The expectation
value of the spin provides an additional vector that may be contracted with the
Levi-Civita tensor, along with $\chi^{\mu\nu}-\chi^{*\mu\nu}$ and $p_{\nu}$. The
details of the dependence on the spin and the imaginary part of $\chi$ are discussed
in the appendix.

The remaining anisotropic term in the spin-summed cross section does not vanish
at this stage. Splitting the $\chi$ terms into
separate spatial and temporal parts and using
$p_{\mu}\cdot p_{\nu}=m_{\pi}^{2}(1-\xi^{2})/2$, the decay rate is proportional to
\begin{eqnarray}
\label{eq-chisplit}
\sum_{s,s'}|{\cal M}|^{2} & \propto & (1+\chi^{00}+\chi^{*00})m_{\pi}^{2}\frac{\xi^{2}
(1-\xi^{2})}{2}-(\chi^{0j}+\chi^{*0j})\hat{p}_{j}m_{\pi}^{2}\frac{\xi^{2}
(1-\xi^{2})}{2} \\
& = & (1+\chi^{0\alpha}+\chi^{*0\alpha})m_{\pi}\xi^{2}(p_{\nu})_{\alpha}.
\end{eqnarray}
This describes the anisotropy of the decay; the rate depends on the alignment of
the outgoing particles with the axis described by $\Re\{\chi^{0j}\}$. However,
when integrated over all angles, the $\hat{p}$ term in (\ref{eq-chisplit}) clearly
gives a vanishing contribution. The net change to the pion decay rate is thus
\begin{equation}
\label{eq-Gamma}
\Gamma=\Gamma_{0}(1+2\chi^{00}),
\end{equation} where $\Gamma_{0}$ is the rate in the
absence of $\chi$.

The contribution of $\chi$ to the modified decay is simpler than the contribution of
a similar coefficient $c_{L}$ in the left-handed fermion sector. The modification
due to $c_{L}$ includes a dependence on the mass ratio $\xi$. However, it is
not surprising that the effects of fermionic Lorentz violation are more intricate,
because $c_{L}$ affects the phase space available to the outgoing fermions in a way
that the weak parameter $\chi$ does not.

\section{Constraints on Weak Sector Lorentz Violation}

\label{sec-constr}

The result (\ref{eq-Gamma}) depends only on the component of $\chi$ that is
isotropic in the pion rest frame. However, pions with different velocities will
see different values of $\chi^{00}$ and will decay at different rates. Thus
there can
actually be substantial anisotropy. Pions with different boosts relative to the
laboratory will decay at different rates, and for a pion beam fixed to the Earth,
the decay rate will exhibit sidereal oscillations.

To study this anisotropy, we must refer to the $\chi$ coefficients in a standard
coordinate frame. The conventional choice for such a frame (in which experimental
constraints on the SME parameters are usually expressed) uses
Sun-centered celestial equatorial coordinates~\cite{ref-bluhm4}.
The $Z$-axis for these coordinates lies parallel to the Earth's axis; the $X$-axis
indicates the direction of the vernal equinox point on the celestial sphere; and the
$Y$-axis is set by the right hand rule. The capital indices $J$ and $K$
used below will refer to
spatial coordinates in this specific frame. The choice of time coordinate is less
essential; there is a standardized choice $T$, but it is usually advantageous to
introduce an offset and use a local time coordinate $T_{\oplus}$, chosen to that
the laboratory $y$-axis and the $Y$-axis lie parallel at $T_{\oplus}=0$.

The isotropic $\chi^{00}$ in the pion rest frame is related to the tensor components
in the Sun-centered frame by a boost. If the pion is moving ultrarelativistically
(with a Lorentz factor $\gamma_{\pi}\gg1$) along the direction $\hat{v}_{\pi}$, the
boosted relationship is
\begin{equation}
\label{eq-chi00}
\chi^{00}=\gamma_{\pi}^{2}\left[\chi_{TT}+\chi_{(TJ)}(\hat{v}_{\pi})_{J}+
\chi_{JK}(\hat{v}_{\pi})_{J}(\hat{v}_{\pi})_{K}\right],
\end{equation}
using the notation $\chi_{(\mu\nu)}=\chi_{\mu\nu}+\chi_{\nu\mu}$.
The size of the effect in the pion rest frame depends on $\gamma_{\pi}^{2}$ and
thus increases quite rapidly with the energy. This serves to enhance the sensitivity
of measurements made with fast-moving pions. Note that, for an
Earthbound laboratory and a beam of relativistic pions, there is no need to
distinguish between the pion velocity relative to the lab and the velocity relative
to the Sun-centered frame; any effects that distinguish between the two are
suppressed by powers of $\gamma_{\pi}$.
At highly relativistic energies, the decay products are also beamed into a narrow
pencil of angles around the direction $\hat{v}_{\pi}$. This effectively washes out
the anisotropy due to terms such as $\chi^{0j}+\chi^{*0j}$.

Searches for evidence of Lorentz-violating neutrino oscillations in the NuMI beam
using the MINOS near detector have already been done~\cite{ref-adamson1,ref-adamson3}. The analyses
looked for sidereal variations in the number of charged current events
in the detector. However, as noted in~\cite{ref-altschul33}, this can alternatively be
interpreted as a search for directional variations in the neutrino beam strength. The
signature of a direction-dependent pion decay rate would be a variation in the number
of charged current events seen in the detector. This idea, along with the sidereal
analysis from~\cite{ref-adamson1}, has already been used to constrain the Lorentz
violating $c_{L}$ coefficients for the second-generation leptons.

The techniques used here shall be quite similar, and much of the analysis is
completely analogous.  However, in both cases, there are additional complications,
and the results should be seen as order of magnitude estimates. The primary concern is
that, in particle physics experiments, the Lorentz violation may affect not just the
rate at which a given interaction or decay occurs, but it may also change the
efficiency with which the outgoing products are detected and identified. These kinds
of detection issues are frequently sidestepped, as in~\cite{ref-abazov}, which
considered Lorentz violation for $t$ quarks but did not address the way the forms of
Lorentz violation involved would also affect the $b$ quark jets used to tag the
$t$ events.

In the process being considered in this paper,
the $\chi$ coefficients will generally modify not just $\Gamma$ but also the
rate at which the NuMI beam neutrinos interact with the detector material.
The charged current interactions that give rise to detectable signals are more
complicated than the simple leptonic decay we have been discussing. However, there
is good reason to believe that the interactions in the detector are less affected by
the Lorentz violation than is the pion decay. The sensitivity of a weak process to
the Lorentz violation coefficients goes as $E^{2}$, where $E$ is the lab frame
energy. Since a decay neutrino carries off, on average, less than half of
each pion's energy, the reactions at the detector are anticipated to be at least 4
time less sensitive to the observable components of $\chi$. Moreover, the effects
of $\chi$ in the detector are generally expected to enhance, not diminish, the final
observable. Because the boosted momenta that the $\chi^{\mu\nu}$ will ultimately
be contracted with tend to point in roughly the same direction in the original pion
decay and in the interaction at the detector,
an enhancement of one reaction will generally be accompanied by an enhancement of the
other as well. The net result is
that the effect of neglecting modifications to the
detector interactions should only be a modest loosening of the bounds that we shall
be able to derive.

Of course, the intensity of the neutrino beam does not provide a direct measurement
of $\Gamma/\Gamma_{0}$ either. What is effectively being measured is actually the
total number of pions that manage to decay in the neutrino beam production region.
The length $D$ of the decay pipe determines the magnitude of the variations in
the neutrino beam strength. For example, if the decay pipe is very long, essentially
all the pions will have time to decay, and the sensitivity to $\Gamma/\Gamma_{0}$
will be lost. Conversely, if the pie is very short, the number of decays will be
directly proportional to $\Gamma/\Gamma_{0}$; however, the beam intensity will be
low, and the detection statistics will suffer accordingly. In general, as $D$
grows, the
fractional sensitivity to $\Gamma/\Gamma_{0}$ decreases, but the statistics
improve.

Measured in the laboratory frame, the pion decay rate is $\Gamma/\gamma_{\pi}$, with
the usual time dilation factor to account for the pions' relativistic motion; this
factor is separate from those appearing in (\ref{eq-chi00}). Of pions moving with
a speed $v_{\pi}\approx1$ along the decay pipe of length $D$, a fraction
$P(D)=1-e^{-\Gamma D/\gamma_{\pi}}$ have sufficient time to decay. Therefore,
a fractional change $\Gamma/\Gamma_{0}$ in the instantaneous decay rate produces
a fractional change in $P(D)$ of
\begin{equation}
\label{eq-pipe}
\left(\left.\frac{1}{P}\frac{dP}{d\Gamma}\right|_{\Gamma=\Gamma_{0}}\right)\Delta
\Gamma=\left(\frac{\Delta \Gamma}{\Gamma_{0}}\right)\frac{\Gamma_{0}D/\gamma_{\pi}}
{e^{\Gamma_{0}D/\gamma_{\pi}}-1}.
\end{equation}
In (\ref{eq-pipe}),  a factor of $\Gamma_{0}/\Gamma_{0}$ has been inserted to make
the expression into a product of two dimensionless quantities.

The decay pipe for the NuMI beam is 677 m long. This length was chosen to be
comparable to the mean decay length for pions with GeV energies, which means the
factor $\Gamma_{0}D/\gamma_{\pi}(e^{\Gamma_{0}D/\gamma_{\pi}}-1)$ appearing in
(\ref{eq-pipe}) will be ${\cal O}(1)$. In fact, for a 6.0-GeV pion (representative
of the peak of the pion distribution), the value of the factor is 0.31.

Following the reasoning from of~\cite{ref-altschul33}, we can determine the
manner in which the dependence of $\Gamma$ on $\chi$ will produce sidereal
variations in the NuMI beam intensity.
The MINOS near detector is located at colatitude $\varphi=42.18^{\circ}$ at Fermilab,
and we will use spherical coordinates $(\theta,\phi)$ describing
the angle between the beam direction and the local zenith direction
($\theta=93.27^{\circ}$) and the azimuthal angle in the plane of the Earth's surface,
measured starting eastward from south ($\phi=203.91^{\circ}$). This makes the beam
direction at a local sidereal time $T_{\oplus}=0$
\begin{eqnarray}
\hat{v}_{\pi} & = & N_{1}\hat{X}+N_{2}\hat{Y}+N_{3}\hat{Z} \\
& = & (\cos\varphi\sin\theta\cos\phi+\sin\varphi\cos\theta)\hat{X}
+(\sin\theta\sin\phi)\hat{Y} \nonumber\\
& & +(-\sin\varphi\sin\theta\cos\phi+\cos\varphi\cos\theta)
\hat{Z} \\
& = & -0.715\hat{X}-0.405\hat{Y}+0.571\hat{Z}.
\end{eqnarray}
The revolution of the beam around the $\hat{Z}$-direction causes the the quantity
$\chi^{00}$ upon with $\Gamma$ depends to vary as
\begin{equation}
\label{eq-chiboost}
\chi^{00}=\gamma_{\pi}^{2}\left[{\cal A}_{0}+{\cal A}_{\omega}\cos(\omega_{\oplus}
T_{\oplus})+{\cal B}_{\omega}\sin(\omega_{\oplus}T_{\oplus})+
{\cal A}_{2\omega}\cos(2\omega_{\oplus}T_{\oplus})+
{\cal B}_{2\omega}\sin(2\omega_{\oplus}T_{\oplus})\right],
\end{equation}
where $\omega_{\oplus}$ is the Earth's sidereal rotation frequency. The Fourier
coefficients appearing in this expression are
\begin{eqnarray}
{\cal A}_{0} & = & \chi_{TT}+N_{3}\chi_{(TZ)}
+N_{3}^{2}\chi_{ZZ}+\frac{1}{2}(1-N_{3}^{2})[\chi_{XX}+\chi_{YY}] \\
{\cal A}_{\omega} & = & N_{1}\chi_{(TX)}+N_{1}N_{3}\chi_{(XZ)}+
N_{2}\chi_{(TY)}+N_{2}N_{3}\chi_{(YZ)} \\
{\cal B}_{\omega} & = & -N_{2}\chi_{(TX)}-N_{2}N_{3}\chi_{(XZ)}+
N_{1}\chi_{(TY)}+N_{1}N_{3}\chi_{(YZ)} \\
{\cal A}_{2\omega} & = & \frac{1}{2}(N_{1}^{2}-N_{2}^{2})\chi_{-}
+N_{1}N_{2}\chi_{(XY)} \\
{\cal B}_{2\omega} & = & -N_{1}N_{2}\chi_{-}
+\frac{1}{2}(N_{1}^{2}-N_{2}^{2})\chi_{(XY)},
\end{eqnarray}
with $\chi_{-}=\chi_{XX}-\chi_{YY}$. (Since all the combinations appearing in the
Fourier coefficients are symmetric, the expressions are all manifestly real.)

The neutrino beam intensity varies with the value of $\chi^{00}$ in the pion frame.
The amplitudes of the beam strength oscillations are given by the ${\cal A}$ and
${\cal B}$ coefficients, times a common sensitivity factor
\begin{equation}
\label{eq-S}
{\cal S}=2\gamma_{\pi}^{2}\left(\frac{\Gamma_{0}D/\gamma_{\pi}}
{e^{\Gamma_{0}D/\gamma_{\pi}}-1}\right)=1.2\times10^{3}.
\end{equation}
The factor of 2 in (\ref{eq-S}) comes directly from the expression for
$\Gamma/\Gamma_{0}$;
the $\gamma_{\pi}^{2}$ (which is the dominant contribution) derives from
(\ref{eq-chiboost}); and, as discussed above, the final factor in parentheses relates
how a variation in $\Gamma/\Gamma_{0}$ affects total number of pion decays.

\begin{table}
\begin{center}
\begin{tabular}{|c|c|}
\hline
Coefficient & Bound \\
\hline
$|\chi_{(TX)}|$ & $6.3\times10^{-5}$ \\
$|\chi_{(TY)}|$ & $6.3\times10^{-5}$ \\
$|\chi_{-}|$ & $1.6\times10^{-4}$ \\
$|\chi_{(XY)}|$ & $1.6\times10^{-4}$ \\
$|\chi_{(XZ)}|$ & $1.1\times10^{-4}$ \\
$|\chi_{(YZ)}|$ & $1.1\times10^{-4}$ \\
\hline
\end{tabular}
\caption{
\label{table-bounds}
Constraints on the magnitudes of the individual $\chi$ coefficients in the
Sun-centered reference frame, assuming only a single coefficient is nonvanishing.}
\end{center}
\end{table}

An analysis of $\chi$ based on the MINOS data mirrors the analyses
from~\cite{ref-adamson1,ref-altschul33}. There is no evidence (at a 3$\sigma$ level)
for any of the oscillation amplitudes in the charged current event rate to be
nonzero. The fractional level of statistical noise in the data was characterized by
a $1\sigma$ dispersion in the power values of $1.8\times10^{-2}$. We therefore
assign $3\sigma$ constraints on the quantities ${\cal SA}$ and
${\cal SB}$ at the $5.4\times10^{-2}$ level.
This corresponds to constraints on the four
separate oscillations amplitudes ${\cal A}_{\omega}$, ${\cal B}_{\omega}$,
${\cal A}_{2\omega}$, and ${\cal B}_{2\omega}$ of $4.5\times10^{-5}$.
By assuming only one components of $\chi$ is nonzero at a time, the constraints
on the ${\cal A}$ and ${\cal B}$ linear combinations can be translated into
bounds on separate coefficients. The results are given in table~\ref{table-bounds}.

\section{Discussion}

\label{sec-disc}

The constraints on the $\chi^{\mu\nu}$ coefficients are at the $10^{-4}$ level,
representing an improvement over those discussed in~\cite{ref-wilschut}, which were
at the $10^{-3}$ level. The bounds cover both $\chi_{(JK)}$ and $\chi_{(TJ)}$
coefficients. However, none of these constraints are particularly tight, compared
with bounds in other sectors of the SME. Moreover, these bounds are not on precisely
the same $\chi$ parameters as might be measured in different processes, because of
the momentum dependence in the $k_{W}$ term of (\ref{eq-chibreakdown}). The present
bounds are also comparable to those obtained on $\chi_{TT}$ in~\cite{ref-nielsen} by
comparing the pion decay rate at different energy scales. Overall,
there is ample room for improvement in this poorly constrained region of parameter space.

The bounds in table~\ref{table-bounds} were derived under the assumption that only a
single component of $\chi$ was nonvanishing. In reality, the MINOS data constrains
the four linear combinations ${\cal A}_{\omega}$, ${\cal B}_{\omega}$,
${\cal A}_{2\omega}$, and ${\cal B}_{2\omega}$. By measuring sidereal changes in
intensity using pion beams with a uniform energy but different terrestrial
orientations, the sensitivity could be expanded to cover the six-parameter subspace
considered spanned by $\chi_{TX}$, $\chi_{TY}$, $\chi_{XY}$, $\chi_{XZ}$, $\chi_{YZ}$,
and $\chi_{-}$. The remaining real coefficients, which determine ${\cal A}_{0}$,
could be constrained by varying the pion energy. Varying the energy only (but not
the beam direction) gives sensitivity to $\chi_{TT}$; this was essentially the
approach taken in~\cite{ref-nielsen}, in which isotropy was assumed. Varying both
direction and energy provides sensitivities to the three remaining quantities
$\chi_{TZ}$, $\chi_{ZZ}$, and $\chi_{XX}+\chi_{YY}$, which describe violations of
isotropy and boost invariance in which the $Z$-axis represents the sole preferred
direction. Finally, using different decaying meson species, with different rest
frame energies, would make it possible to disentangle the $k_{W}$ coefficients
from the others than make up the effecive quantity $\chi$.

There has actually been some confusion about how well the $k_{\phi\phi}$,
$k_{\phi W}$, and $k_{W}$ that make up $\chi$ are known.
In~\cite{ref-anderson}, much stronger bounds on some of the same SME coefficients
were quoted. However, those bounds are not really accurate.
There is a particular class of CPT-even SME coefficients---the $c$ coefficients for
fermions, $k_{\phi\phi}$ in the Higgs sector, subsets of the $k_{F}$ and $k_{W}$
for electromagnetic and weak fields---that are observable only as differences
between different sectors. Coordinate redefinitions can actually move the
coefficients from one sector to another, but the differences between sectors are
invariant under such transformations~\cite{ref-altschul9}.
In some cases, such as analyses of the thresholds for
ordinarily forbidden electron-photon processes like $\gamma\rightarrow e^{-}+e^{+}$,
it is easy to identify unambiguously the specific difference involved. The
threshold for the photon decay depends on the difference between photon-sector and
electron-sector coefficients, because a comparison is effectively being made between
the electron and photon dispersion relations.

However, in many low-energy experiments, precisely which sectors are being compared
is not clear. In atomic clock experiments, the coefficients that are actively being
studied are really being compared with some sort of aggregate coefficients for bulk
matter. These aggregate coefficients are linear combinations of the coefficients for
photons, electrons, protons, and neutrons (the latter two themselves being
combinations of gluon and quark coefficients). In many cases, the comparison of
different sectors is
left implicit; it is assumed that there is only Lorentz violation for a single
standard model field, and bounds are quoted for the corresponding coefficients in
that sector.

However, only the coefficients for
those sectors that are actually involved in a given process can be constrained by
studying that process. In~\cite{ref-anderson}, the authors took constraints on
various coefficients (which were not quoted as differences) as if they represented
absolute bounds on the coefficients in a single sector. Then they used coordinate
transformations to move the Lorentz violation to the weak sector and reinterpreted
the experimental results as constraints on weak-sector Lorentz violation.
Unfortunately, this methodology is not correct. It must be remembered that the
experimental results really constrain only differences between coefficients in
various sectors, and that the weak sector is not included as part of any of those
differences. Since the differences between sectors are
invariant under the redefinitions used in~\cite{ref-anderson}, the
physical observables cannot be interpreted as depending on weak-sector SME
coefficients.

The constraints in~\cite{ref-anderson} on these CPT-even coefficients are not
wholy meaningless, however. The weak sector does affect the physics of
ordinary matter, through electroweak mixing and radiative mixing. The
bounds quoted in~\cite{ref-anderson} actually provide rough estimates of how large
Lorentz violation in the weak sector can be without additional fine tuning. If
Lorentz violation in the weak sector were much larger, there would need to be
unnatural cancellations to prevent the large Lorentz violation from being
transmitted to other sectors. However, naturalness, or the absence of fine tuning,
is as aesthetic condition, not a rigorous one.

Considerations of naturalness and direct constraints via measurement
therefore provide complementary
approaches. Direct bounds on Lorentz violation, such as those
discussed in this paper, are more rigorous. This makes them an important part of
the developing analysis of the SME---espcially in the weak sector, which is, so far,
quite poorly constrained.

\section*{Acknowledgements}

The author is grateful to  J. P. Noordmans,
R. Timmermans, K. K. Vos, and H. W. Wilschut for helpful discussions.

\appendix

\section*{Appendix: Dependence on the Imaginary Part of $\chi$}

We return now to the theoretical question of how spin measurements can
add sensitivity to the imaginary part of the $\chi$ tensor describing
weak and Higgs sector Lorentz violations.  We shall replace the spin summed
$u\bar{u}$ in (\ref{eq-M2}) with the expression corresponding to specific
outgoing spin state.

Throughout this paper, we have assumed that the inital pion was negatively
charged, so that the daughter particles were $\mu^{-}$ and $\bar{\nu}_{\mu}$.
This choice was made simply for definiteness, and the result for the total
decay rate $\Gamma$ is not affected by the choice. This is the case because the
real part of $\chi$ is invariant under charge conjugation (C). However, the
imaginary part changes sign under C, so the charges of the particles involved
in the process are significant. In this appendix, we shall continue to assume
that the parent particle is a $\pi^{-}$; however, if it were a $\pi^{+}$, all
the effects of the imaginary part of $\chi$ would have their signs reversed.

Since the chirality matrix $\gamma_{5}$ plays a prominent role in the
matrix element, it is convenient to use the Weyl representation of the Dirac
matrices. With relativistic normalization, the spinor for a fermion of momentum
$\vec{p}_{\mu}=p_{\mu 3}\hat{z}$ is
\begin{equation}
u(p_{\mu})=\left[
\begin{array}{c}
\sqrt{E_{\mu}-p_{\mu 3}\sigma_{3}}\xi \\
\sqrt{E_{\mu}+p_{\mu 3}\sigma_{3}}\xi
\end{array}
\right]=\left[
\begin{array}{c}
\sqrt{p_{\mu}\cdot\sigma}\xi \\
\sqrt{p_{\mu}\cdot\bar{\sigma}}\xi
\end{array}
\right],
\end{equation}
where $\sigma=(1,\vec{\sigma})$ and $\bar{\sigma}=(1,-\vec{\sigma})$.
The two-spinor $\xi=[\xi_{1},\xi_{2}]^{t}$ determines the spin state.

The quantity that appears in $|{\cal M}|^{2}$ is
\begin{equation}
u\bar{u}=uu^{\dag}\gamma_{0}=\left[
\begin{array}{cc}
\sqrt{p_{\mu}\cdot\sigma}\xi\xi^{\dag}\sqrt{p_{\mu}\cdot\bar{\sigma}} &
\sqrt{p_{\mu}\cdot\sigma}\xi\xi^{\dag}\sqrt{p_{\mu}\cdot\sigma} \\
\sqrt{p_{\mu}\cdot\bar{\sigma}}\xi\xi^{\dag}\sqrt{p_{\mu}\cdot\bar{\sigma}} &
\sqrt{p_{\mu}\cdot\bar{\sigma}}\xi\xi^{\dag}\sqrt{p_{\mu}\cdot\sigma}
\end{array}
\right].
\end{equation}
This can be expanded in terms of the explicit spinor $\xi$ and momentum
$p_{\mu}$. For example, the upper left block is
\begin{eqnarray}
\sqrt{p_{\mu}\cdot\sigma}\xi\xi^{\dag}\sqrt{p_{\mu}\cdot\bar{\sigma}}
& = &\left[
\begin{array}{cc}
\sqrt{E_{\mu}-p_{\mu3}}\xi_{1}\xi_{1}^{*}\sqrt{E_{\mu}+p_{\mu3}} &
\sqrt{E_{\mu}-p_{\mu3}}\xi_{1}\xi_{2}^{*}\sqrt{E_{\mu}-p_{\mu3}} \\
\sqrt{E_{\mu}+p_{\mu3}}\xi_{2}\xi_{1}^{*}\sqrt{E_{\mu}+p_{\mu3}} &
\sqrt{E_{\mu}+p_{\mu3}}\xi_{2}\xi_{2}^{*}\sqrt{E_{\mu}-p_{\mu3}}
\end{array}
\right] \\
& = & \left[
\begin{array}{cc}
m_{\mu}\xi_{1}\xi_{1}^{*} & (E_{\mu}-p_{\mu3})\xi_{1}\xi_{2}^{*} \\
(E_{\mu}+p_{\mu3})\xi_{2}\xi_{1}^{*} & m_{\mu}\xi_{2}\xi_{2}^{*}
\end{array}
\right].
\end{eqnarray}
The other blocks reduce in a similar fashion.


It is also necessary to cast the quantities $\xi_{j}\xi_{k}^{*}$ in
a more convenient form. The two-spinor product $\xi\xi^{\dag}$ may be concisely
expressed in terms of the expectation value $\langle\vec{\sigma}\rangle=
\xi^{\dag}\vec{\sigma}\xi$ of the Pauli spin vector in the two-dimensional
spinor space. It is important, however, that this is not the same as the
expectation value of the full muon spin
$\langle\vec{\Sigma}\rangle=(2E_{\mu})^{-1}u^{\dag}\vec{\Sigma}u$. For the
compoment of the spin along the direction of the motion,
$\langle\Sigma_{3}\rangle=\langle\sigma_{3}\rangle$, since the helicity
component of the spin commutes with a boost along the momentum
direction. However, the expectation
values of the transverse spin components are suppressed by relativistic effects,
so that $\langle\Sigma_{j}\rangle=\frac{m_{\mu}}{E_{\mu}}
\langle\sigma_{j}\rangle$ for $j=1$ or 2. We shall return to this fact
presently. However, for the moment we shall simply make use of the expression
\begin{equation}
\xi\xi^{\dag}=\left[
\begin{array}{cc}
\xi_{1}\xi_{1}^{*} & \xi_{1}\xi_{2}^{*} \\
\xi_{2}\xi_{1}^{*} & \xi_{2}\xi_{2}^{*}
\end{array}
\right]=\frac{1}{2}\left[
\begin{array}{cc}
1+\langle\sigma_{3}\rangle & \langle\sigma_{-}\rangle \\
\langle\sigma_{+}\rangle & 1-\langle\sigma_{3}\rangle
\end{array}
\right],
\end{equation}
with $\sigma_{\pm}=\sigma_{1}\pm\sigma_{2}$.

The full $4\times4$ matrix $u\bar{u}$ is still rather cumbersome, and it can be
expanded using the sixteen Dirac matrices as a basis. However,
when this expression replaces $(\!\not\!p_{\mu}+m_{\mu})$ in the trace (\ref{eq-M2}),
only those
terms that are constructed from odd numbers of $\gamma$ matrices can contribute.
This means the terms proportional to $\gamma_{\alpha}$ or
$\gamma_{5}\gamma_{\alpha}$; equivalently, the only contributions come from
the portion of $u\bar{u}$ that is block off-diagonal. (In fact, because of
the way $u\bar{u}$ is sandwiched by chiral projectors, only a single
off-diagonal block will ultimately contribute; however, it is computationally
simpler to keep both off-diagonal blocks.)
The block off-diagonal portion is
given by (dropping the subscripts from $m_{\mu}$, $E_{\mu}$, and $p_{\mu3}$ for
brevity)
\begin{equation}
u\bar{u}=\frac{1}{2}\left[
\begin{array}{cc}
\ldots &
\begin{array}{cc}
(E-p)(1+\langle\sigma_{3}\rangle) & m\langle\sigma_{-}\rangle \\
m\langle\sigma_{+}\rangle & (E+p)(1-\langle\sigma_{3}\rangle)
\end{array}
\\
\begin{array}{cc}
(E+p)(1+\langle\sigma_{3}\rangle) & m\langle\sigma_{-}\rangle \\
m\langle\sigma_{+}\rangle & (E-p)(1-\langle\sigma_{3}\rangle)
\end{array}
& \ldots
\end{array}
\right].
\end{equation}
The expansion of this in terms of Dirac matrices is now straightforward.
Leaving off the terms that will not contribute, we have
\begin{equation}
\label{eq-uubarspin}
u\bar{u}=\frac{1}{2}\left(E_{\mu}\gamma_{0}-p_{\mu3}\gamma_{3}
+p_{\mu3}\langle\sigma_{3}\rangle\gamma_{5}\gamma_{0}
-E_{\mu}\langle\sigma_{3}\rangle\gamma_{5}\gamma_{3}
-m_{\mu}\langle\sigma_{1}\rangle\gamma_{5}\gamma_{1}
-m_{\mu}\langle\sigma_{2}\rangle\gamma_{5}\gamma_{2}\right)+\ldots.
\end{equation}
Note that the spin-independent term in (\ref{eq-uubarspin}) is simply
the $\!\not\!p_{\mu}$ that was already present in (\ref{eq-M2}). However,
there is also a new vector contracted with $\gamma_{5}\gamma_{\alpha}$.
With $w=(\vec{p}_{\mu}\cdot\langle\vec{\Sigma}\rangle,E_{\mu}\langle
\vec{\Sigma}\rangle)$, we have
\begin{equation}
u\bar{u}=\frac{1}{2}\left(\!\not\!p_{\mu}+w^{\alpha}\gamma_{5}
\gamma_{\alpha}\right)+\ldots.
\end{equation}

So the analogue of (\ref{eq-M2simp}) without the spin sum is
\begin{equation}
|{\cal M}|^{2}\propto
(g^{0\beta}+\chi^{0\beta})(g^{0\gamma}+\chi^{*0\gamma}){\rm tr}\left\{
\!\not\!p_{\nu}\gamma_{\beta}(\!\not\!p_{\mu}-\!\not\!w)\gamma_{\gamma}-
\!\not\!p_{\nu}\gamma_{\beta}\gamma_{5}(\!\not\!p_{\mu}-\!\not\!w)
\gamma_{\gamma}\right\}.
\end{equation}
Whatever basis is chosen, the two muon spin states have opposite values of
$\langle\vec{\Sigma}\rangle$, so summing over both of them sends the
$w$-dependent terms to 0. For a nonrelativistic muon, there are substantial
contributions from both helicity states; however, if the muon is
ultrarelativistic ($m_{\mu}\rightarrow0$), the key quantity $p_{\mu}-w$
vanishes for the positive helicity states, and there are (as expected) no
contributions from right-handed muons.

The $w$ vector influences both the $\chi^{\mu\nu}+\chi^{*\mu\nu}$ and
$\chi^{\mu\nu}-\chi^{*\mu\nu}$ terms in $|{\cal M}|^{2}$. However, our main
purpose here has been to obtain the dependence on the latter, since the
imaginary part of $\chi$ is detectable only through spin correlations. The
relevant term in the spin-dependent generalization of
(\ref{eq-M2final}) is
\begin{equation}
\label{eq-chiIm}
4ig^{0\gamma}(\chi^{0\beta}-\chi^{*0\beta})\epsilon_{\alpha\beta\delta\gamma}
(p_{\nu})^{\alpha}(-w^{\delta})=8E_{\mu}\vec{\chi}\cdot(\vec{p}_{\nu}\times
\langle\vec{\Sigma}\rangle),
\end{equation}
where $\chi_{j}=(\chi^{0j}-\chi^{*0j})/2i$.
Since the kinematics are not affected by the muon spin in any way, the
anisotropic structure in (\ref{eq-chiIm})
carries through to appear in the final spin-dependent differential
decay rate. Only the spin components orthogonal to the momentum
$\vec{p}_{\nu}=-\vec{p}_{\mu}$ contribute to the expression. Therefore, the
effect does not grow with energy; the factor of $E_{\mu}$ only serves to
balance the relativistic suppresion of the transverse part of
$\langle\vec{\Sigma}\rangle$.

Measuring this term would be extremely challenging
experimentally, but its existence is nonetheless theoretically interesting,
and we have definitively demonstrated that the imaginary part of $\chi$ is,
in principle, an observable quantity. This is in constrast with some other
SME parameters, which may naively appear to be physically significant but,
in fact, cannot be observed at all.

\end{document}